\documentclass[12pt,a4paper]{article}
\usepackage{url}
\usepackage{jcappub}
\usepackage{amsfonts}
\usepackage{subfigure}
\usepackage{graphicx}
\usepackage{appendix}

\def\eV{{\rm\thinspace eV}}
\def\g{{\rm\thinspace g}}

\def\h18{\hbox{H1821$+$643\,}}

\newcommand{\beq}{\begin{equation}}
\newcommand{\be}{\begin{equation}}
\newcommand{\ee}{\end{equation}}
\newcommand{\bea}{\begin{eqnarray}}
\newcommand{\eea}{\end{eqnarray}}
\newcommand{\eeq}{\end{equation}}
\newcommand{\bef}{\begin{figure}}
\newcommand{\eef}{\end{figure}}

\newcommand{\bP}{{\bf P}}
\newcommand{\GeV}{{\rm GeV}}

\newcommand{\bB}{{\bf B}}

\def\g{\ensuremath{g_{a\gamma}}}

\begin{document}

\title{On the applicability of the Landau-Zener formula \\ to axion-photon conversion}

\author{Pierluca Carenza,}
\author{M.C.~David Marsh}

\affiliation{The Oskar Klein Centre, Department of Physics, Stockholm University, Stockholm 106 91, Sweden}

\emailAdd{pierluca.carenza@fysik.su.se} \emailAdd{david.marsh@fysik.su.se}

\abstract{Axions and photons resonantly interconvert in regions where the plasma frequency approximately equals the axion mass. This process is directly analogous to an avoided level crossing in quantum mechanics, for which the celebrated Landau-Zener (LZ) formula provides a simple, non-perturbative solution for the conversion probability. The LZ formula is commonly used in studies of axion-photon conversion; however, in this context, it relies on the assumption that the magnetic field variation is small compared to variations of the plasma frequency, which is frequently not the case in real plasmas.
We derive a generalised version of the LZ formula by allowing the boundaries to be located at a scale that is similar to the inhomogeneities. We find that the LZ formula fails when the oscillation range is small compared to the resonance region. This failure is more severe in the adiabatic limit, when the plasma frequency varies slowly, resulting in
a conversion probability that is not maximal as opposed to a naive application of the LZ formula.
Moreover, we consider circumstances where the generalised LZ formula does not apply and present an alternative  semi-classical approximation with complementary regime of validity.
}
\maketitle

\section{Introduction}\label{sec:intro}

The strong CP problem is an unsolved problem of the Standard Model (SM) and one of the most elegant solutions is given by the Peccei-Quinn mechanism with the introduction of a hypothetical pseudoscalar particle: the axion~\cite{Peccei:1977ur,Peccei:1977hh,Weinberg:1977ma,Wilczek:1977pj}.
A general prediction of many axion models is an axion-photon interaction described by the following Lagrangian
\begin{equation}
    \mathcal{L}=-\frac{1}{4}g_{a\gamma}a F_{\mu\nu}\tilde{F}^{\mu\nu}\;,
    \label{eq:lag}
\end{equation}
where $a$ is the axion field, $F_{\mu\nu}$ is the electromagnetic tensor, $\tilde{F}_{\mu \nu}$ its dual and $g_{a\gamma}$ is the axion-photon coupling. In a more general framework, many Grand Unified Theories and models from string theory involve pseudoscalar particles coupled with photons through eq.~\eqref{eq:lag}~\cite{Green:1987sp, Svrcek:2006yi, Halverson:2019cmy, Demirtas:2021gsq}. Because of their similarity with axions, these particles are dubbed Axion-Like Particles (ALPs). In this work, for simplicity, we use the term `axion' to refer to both axions and ALPs.

The axion-photon interaction leads to a rich phenomenology and important observational consequences. In particular, an axion in an external magnetic field can convert into a photon~\cite{Raffelt:1987im} and this phenomenon is used to produce and directly detect axions in laboratory experiments~\cite{Anselm:1986gz,VanBibber:1987rq,Ehret:2010mh,Bahre:2013ywa}, reveal axions that constitute the dark matter~\cite{Sikivie:1983ip,Asztalos:2003px,Brun:2019lyf} or probe axions produced in the Sun~\cite{Sikivie:1983ip,vanBibber:1988ge,Armengaud:2019uso}.
The phenomenological relevance of axion-photon conversions stimulates a better theoretical understanding of this phenomenon, which is analogous to the well known quantum mechanical evolution of a two-level system~\cite{Raffelt:1987im}. 

In this work we focus on so-called `resonant' axion-photon mixing, which occurs in regions where the plasma frequency matches the axion mass and the conversion is more efficient.\footnote{By contrast, `non-resonant' mixing is not localised but generated from contributions over the axion trajectory and is, in the perturbative limit, only sensitive to a single Fourier mode of $B/\omega_{\rm pl}^2$ \cite{Marsh:2021ajy}.} This phenomenon is relevant, for example, in the case of dark matter axions converting into radio waves in the magnetosphere of a neutron star~\cite{Pshirkov:2007st}; to conversions in the primordial magnetic field of the early Universe, leading to distortions of the cosmic microwave background blackbody spectrum~\cite{Mirizzi:2009nq,Mukherjee:2018oeb} and for spectral modifications of photon spectra from active galactic nuclei~\cite{Hochmuth:2007hk}.
If the axion-photon conversion is mapped into the language of a quantum two-level system, the resonant conversion corresponds to the well-known avoided level crossing phenomenon~\cite{Landau:1991wop}. The Landau-Zener formula for the conversion probability \cite{Landau:1932,Majorana:1932,Stueckelberg:1932,Zener:1932ws} can then be directly applied to the problem of axion-photon mixing.\footnote{The same formalism is important in neutrino physics where it describes why neutrinos with a small mixing angle can achieve a large conversion probability in a medium with non-constant electron density due to resonant conversions (Mikheyev-Smirnov-Wolfenstein effect)~\cite{Wolfenstein:1977ue,Mikheyev:1985zog,Bethe:1986ej}. In the neutrino case the LZ formula can be applied when the resonance occurs on an oscillation length $l_{\rm osc}$ small compared with the scale over which the electron density varies. This condition is typically satisfied in applications.} 

However, this approach has a limited applicability. Indeed, a strong assumption behind the LZ formula in this context is that the axion-photon mixing takes place on a length scale that is sufficiently small compared to the scales associated to the magnetic field and plasma inhomogeneities. In realistic astrophysical environments, however, this hierarchy of scales is not always realised.

For the axion-photon system we identify three relevant length scales: two of them are connected with the magnetic field ($l_{\rm B}$) and plasma frequency ($l_{\rm pl}$) variations and the last one characterises the length-scale for axion-photon mixing ($l_{\rm osc}$). The validity of the LZ formula is limited when $l_{\rm osc}\ll l_{\rm pl}\ll l_{\rm B}$. This condition is very restrictive and often violated in realistic plasmas when there is a correlation between the magnetic field and plasma fluctuations, i.e.~$l_{\rm pl}\sim l_{\rm B}$. 

The LZ formula is derived under the assumptions of a constant, non-vanishing magnetic field, a linearly increasing plasma frequency, and asymptotically imposed boundary conditions. To quantify the impact of finite-distance deviations from these assumptions, we derive a generalised version of the formula that allows for boundary conditions to be imposed a finite distance from the crossing point (still considering a constant magnetic field). We find that the conversion probabilities become oscillatory along the trajectory and that the deviations from the LZ formula become particularly pronounced for strong (i.e.~`adiabatic') mixing  and when the axion-photon oscillation length becomes comparable to the scale of the variation of the plasma frequency, i.e.~$l_{\rm osc}\sim l_{\rm  pl}\ll l_{\rm B}$. We then point out that axion-photon mixing is analogous to spin-precession in a fictitious, time-dependent magnetic field (cf.~\cite{Werner1984,Kim:1987ss} for a similar discussion in the quantum mechanical context). We also discuss the regime of validity of the LZ formula in the light of our generalised result, and note that the asymptotic convergence to the LZ formula is rather slow. Moreover, we consider arbitrary, but slowly varying, magnetic fields and plasma frequencies and use the Wentzel–Kramers–Brillouin (WKB) approximation to derive an analytic, approximate expression for the conversion probability. This formula applies, in particular, to the case  $l_{\rm osc}\sim l_{\rm pl}\sim l_{\rm B}$.

\section{The Landau-Zener formula}\label{sec:LZ}

Energy level crossing is a topic of great interest in Quantum Mechanics (QM) in connection, for instance, to atomic physics~\cite{Zener:1932ws,Landau:1932}, ultracold molecules~\cite{2008NatPh...4..223L} or to give insights on quantum phase transitions~\cite{Zurek:2005kod} (see also~\cite{Ivakhnenko:2022sfl,Kofman:2022nxo}). The simplest model with two energy levels approaching each other and avoiding the crossing because of their interaction was studied by Landau~\cite{Landau:1932} and Zener~\cite{Zener:1932ws} in early days of QM. In this section we review ref.~\cite{Zener:1932ws} in order to establish notation and analytic results relevant for our subsequent extensions. 

We consider a two-level system evolving in time as prescribed by the time-dependent Hamiltonian $H_{0}(t)$. The normalised eigenstates of $H_{0}(t)$ are $\psi_{1}$ and $\psi_{2}$ with energy eigenvalues $E_{1}(t)$ and $E_{2}(t)$, respectively. The crossing point is the time, conventionally set to zero, $t_{c}=0$, where the energy eigenvalues are degenerate $E_{1}(0)=E_{2}(0)$. If a perturbation $V$ is added such that the total Hamiltonian, $H=H_{0}+V$, the level crossing is avoided. The interaction $V$ is assumed to be non-negligible only for a finite period around $t_c$, so that a system initially (at the time $t_{i}\to-\infty$) prepared in the state $\psi_{2}(t_{i})$ evolves, after the avoided crossing at $t=0$, into a final state (at the time $t\to\infty$)
\begin{equation}
    \psi=\alpha \psi_{1}+\beta \psi_{2}\;,
\end{equation}
where $\alpha$ and $\beta$ are complex constants (satisfying $|\alpha|^2 + |\beta|^2=1$). The probability of observing the system in the asymptotic energy eigenstate $\psi_1$ is clearly $|\alpha|^2$. The Landau-Zener (LZ) formula is a celebrated non-perturbative solution for this transition probability, as we now review.

We consider the  Hamiltonian
\begin{equation}
H=\left(\begin{array}{cc}
\varepsilon_{1}(t)&\varepsilon_{12}^{*}(t)\\
\varepsilon_{12}(t)&\varepsilon_{2}(t)\end{array}\right)\;,
    \label{eq:HQM}
\end{equation}
where $\varepsilon_{i}$ for $i=1,2$ are real functions and $\varepsilon_{12}$ is a complex function. Two simplifying assumptions are required to derive the LZ formula: first, the resonance happens for a time short enough that $ \Delta \varepsilon$ can be approximated as linear in time around $t=0$, i.e.~
\begin{equation}
 \Delta \varepsilon=\varepsilon_{1}-\varepsilon_{2}=\varepsilon_{0}\left(\frac{t}{t_{0}}\right)\;;
\label{eq:delta2}
\end{equation}
second, the mixing parameter $\varepsilon_{12}$ changes slowly compared to the energy difference of the two states $\Delta \varepsilon$, and can be approximated as constant.

\begin{figure}[t!]
	\centering
		\vspace{0.cm}
		\includegraphics[width=0.6\columnwidth]{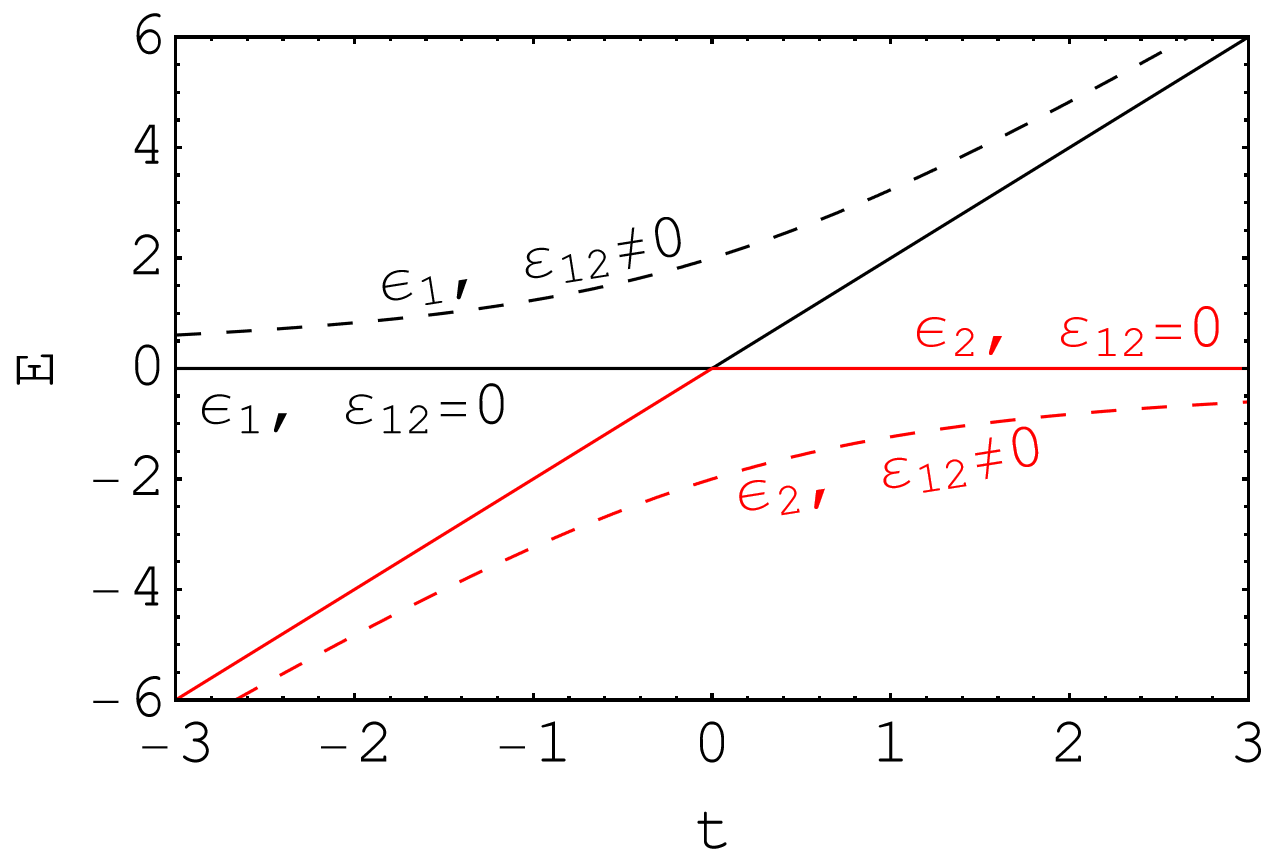}
		\caption{A schematic picture of the level repulsion for the two eigenvalues of the Hamiltonian in eq.~\eqref{eq:HQM} in absence (solid lines) and presence (dashed lines) of an off diagonal term $\varepsilon_{12}$.} 
		\label{fig:levelrepulsion}
	\end{figure}

This simple system highlights some of the most important properties of a two-level system in presence of an avoided level crossing~\cite{Landau:1991wop}. In absence of interactions, $\varepsilon_{12}=0$, the eigenvalues of the Hamiltonian are precisely given by its diagonal components, $\varepsilon_{1}$ and $\varepsilon_{2}$, which cross  at $t_{c}=0$. This situation is shown in fig.~\ref{fig:levelrepulsion}, where the energy eigenvalues for $\varepsilon_{12}=0$ are indicated by solid lines as function of the time, where different colours corresponds to different eigenvalues. The intersection between the two lines at the resonance point is absent if $\varepsilon_{12}\neq0$ (the dashed lines in fig.~\ref{fig:levelrepulsion}). This phenomenon, known as level repulsion or avoided level crossing, is a very general result~\cite{Landau:1991wop,vonNeuman}. 

The LZ formula~\cite{Landau:1932,Majorana:1932,Stueckelberg:1932,Zener:1932ws} is derived by calculating the asymptotic transition probability going from $\psi_2$ at $t\to - \infty$ to $\psi_1$ at $t\to \infty$. The Schrödinger equation for the Hamiltonian in eq.~\eqref{eq:HQM} gives
\begin{equation}
    \begin{split}
        i\dot{\psi}_{1}&=e^{-i\int dt\,\Delta \varepsilon}\varepsilon_{12}\psi_{2}\,,\\
        i\dot{\psi}_{2}&=e^{i\int dt\,\Delta \varepsilon}\varepsilon_{12}^{*}\psi_{1}\;,\\
    \end{split}
    \label{eq:schr}
\end{equation}
and the initial conditions are given by
\begin{equation}
    \psi_{1}(t=-\infty)=0\quad |\psi_{2}(t=-\infty)|=1\;.
\end{equation}
The two first-order equations in eq.~\eqref{eq:schr} can be combined into a single second-order equation
\begin{equation}
\frac{d^{2}\Psi_{1}}{dw^{2}}+\left(1+\frac{1}{2n}-\frac{w^{2}}{4n^{2}}\right)\Psi_{1}=0\;,
\label{eq:final1}
\end{equation}
in terms of the time new variable $w=\sqrt{\frac{\varepsilon_{0}}{t_{0}}}t$, where $\Psi_{1}=e^{\frac{i}{2}\int dt\,\Delta \varepsilon}\psi_{1}$, we define the adiabaticity parameter  
\begin{equation}
    \gamma=t_{0}\frac{|\varepsilon_{12}|^{2}}{\varepsilon_{0}}\,,
\end{equation}
and $n=i\gamma$. 
The general solution of eq.~\eqref{eq:final1} is expressed in terms of the parabolic cylinder functions~\cite{Abramowitz} and undetermined, complex integration constants $c_{1},\, c_2$  (cf.~\cite{vitanov1996landau} for a similar solution and a more detailed discussion)
\begin{equation}
    \Psi_{1}(w)=c_{1}\, D_{-1-n}\left(\frac{iw}{\sqrt{n}}\right)+c_{2}\, D_{n}\left(\frac{w}{\sqrt{n}}\right)\;.
\label{eq:Dsol1}
\end{equation}
The parabolic cylinder functions have the following asymptotics~\cite{Whittaker}
\begin{equation}
    \begin{split}
        &D_{-1-n}\left(e^{i\frac{\pi}{4}}\frac{w}{\sqrt{\gamma}}\right)\xrightarrow[w\to-\infty]{}e^{-i\frac{w^{2}}{4\gamma}}e^{-i\frac{\pi}{4}}e^{\frac{\pi}{4}\gamma}\left(\frac{w}{\sqrt{\gamma}}\right)^{-1-i\gamma}\;,\\
        &D_{n}\left(e^{i\frac{\pi}{4}}\frac{w}{\sqrt{\gamma}}\right)\xrightarrow[w\to-\infty]{}e^{-i\frac{w^{2}}{4\gamma}}e^{-\frac{\pi}{4}\gamma}\left(\frac{w}{\sqrt{\gamma}}\right)^{i\gamma}\;.
    \end{split}
    \label{eq:Dand}
\end{equation}
The initial condition $\lim_{t\to - \infty} \psi_1=0$ (or, equivalently, $\lim_{w\to-\infty}\Psi(w)=0$) translates into $c_{2}=0$. The second initial condition stipulates  $\lim_{t\to - \infty}|\psi_2|=1$, however, directly enforcing this condition is complicated by the oscillatory nature of the general solution. Following ref.~\cite{Zener:1932ws}, we note that the remaining integration constant, $c_1$, can be determined by requiring that the maximum oscillation probability at $w\to \infty$, which is reached for $\gamma\to\infty$, saturates to one. The relevant future-asymptotic behaviour of the parabolic cylinder function is 
\begin{equation}
\begin{split}
D_{-1-n}&\left(e^{i\frac{\pi}{4}}\frac{w}{\sqrt{\gamma}}\right)\xrightarrow[w\to\infty]{}e^{-i\frac{w^{2}}{4\gamma}}e^{-i\frac{\pi}{4}}e^{\frac{\pi}{4}\gamma}\left(\frac{w}{\sqrt{\gamma}}\right)^{-1-i\gamma}+\\
&+\frac{\sqrt{2\pi}}{\Gamma(1+i\gamma)}e^{i\frac{w^{2}}{4\gamma}}e^{-\frac{\pi}{4}\gamma}\left(\frac{w}{\sqrt{\gamma}}\right)^{i\gamma}\;,\\
\end{split}
\label{eq:infbound}
\end{equation}
and the initial condition gives $c_{1}=\sqrt{\gamma}e^{-\frac{\pi}{4}\gamma}$. The final solution is then given by
\begin{equation}
    \Psi_{1}(w)=\sqrt{\gamma}e^{-\frac{\pi}{4}\gamma}D_{-1-n}\left(\frac{iw}{\sqrt{n}}\right)\; ,
    \label{eq:LZsol}
\end{equation}
for $-\infty<w<\infty$. 
The asymptotic  transition probability follows immediately from eq.~\eqref{eq:LZsol}, and it is given by
\begin{equation}
\begin{split}
    P&=\lim_{w\to\infty}|\Psi_{1}(w)|^{2}=
    \lim_{w\to\infty} \gamma e^{-\frac{\pi}{2}\gamma} \Big|D_{-1-n}\left(\frac{iw}{\sqrt{n}}\right)\Big|^2
    =\\
    &= \lim_{w\to\infty} \gamma e^{-\frac{\pi}{2}\gamma}\left[\left(\frac{w}{\sqrt{\gamma}}\right)^{-2}e^{\frac{\pi}{2}\gamma}+\frac{2\pi}{|\Gamma(1+i\gamma)|^{2}}e^{-\frac{\pi}{2}\gamma}+2\sqrt{2\pi}\operatorname{Re}\left(\frac{e^{-i\frac{w^{2}}{2\gamma}}e^{-i\frac{\pi}{4}}}{\Gamma(1-i\gamma)}\left(\frac{w}{\sqrt{\gamma}}\right)^{-1-2i\gamma}\right)\right]=\\
    &=
\frac{2\pi \gamma}{|\Gamma(1+i\gamma)|^2}e^{-\pi \gamma} =    
    1-e^{-2\pi\gamma}\;,
\end{split}
    \label{eq:LZ1}
\end{equation}
where $\operatorname{Re}$ denotes the real part and, in the last step, we have used that 
\begin{equation}
    |\Gamma(1+i\gamma)|^{2}=\frac{\pi \gamma}{\sinh(\pi\gamma)}=\frac{2\pi \gamma}{e^{\pi\gamma}-e^{-\pi\gamma}}\;.
\end{equation}
The same result was recently derived with contour integration methods~\cite{Wittig:2005}, but the proof presented here and discussed in ref.~\cite{Zener:1932ws} is easily generalised to different boundary conditions, as we now discuss.

\section{The Landau-Zener formula for the axion-photon system}
\label{sec:LZaxion}
In this section, we apply the formalism developed section~\ref{sec:LZ} to the axion-photon system, and generalise the solution to  initial conditions placed a finite distance from the level crossing. We show how this results in new characteristic properties of the solutions.

In the most general case, the axion-photon system is composed of three levels: the parallel $A_{\parallel}$ and perpendicular $A_{\perp}$ components of the vector potential (with respect to the axion and photon momentum), and the axion $a$~\cite{Raffelt:1987im}. The axion-photon interaction is described by the following Lagrangian~\cite{Raffelt:1987im}
\begin{equation}
\mathcal{L}=-\frac{1}{4}F^{\mu\nu}F_{\mu\nu}+\frac{1}{2}\left(\partial_{\mu}\,a\partial^{\mu}a-m_{a}^{2}a^{2}\right)-\frac{g_{a\gamma}}{4}a\,F_{\mu\nu}\tilde{F}^{\mu\nu}\, , 
\end{equation}
where $a$ is the axion field, $A_{\mu}$ is the photon field, $F_{\mu\nu}$ is the electromagnetic tensor.

In the following, we neglect the effects of Faraday rotations and QED birefringence, which  are negligible in many applications.
The linearised equations of motion in an environment with plasma frequency $\omega_{\rm pl}$ and external, static magnetic field $\bB_{0}$ are given by
\begin{equation}
\begin{split}
(\Box + m_a^2)a &=- \g \dot {\bf A} \cdot {\bf B}_0 \, ,\\
(\Box + \omega_{\rm pl}^2) {\bf A} &= \g \dot a\, {\bf B}_0 \, . 
\end{split}
\label{eq:eom1}
\end{equation}
We consider, conventionally, the axion and photon momentum to be directed along the $z$-axis. When the axions are relativistic the differential equations effectively reduce to first order  since $\omega^{2}+\partial_{z}^{2}= (\omega+i \partial_{z})(\omega-i \partial_{z})\simeq2\omega(\omega-i \partial_{z})$, where $\omega$ is the axion (and photon) energy. With these simplifications, eq.~\eqref{eq:eom1} becomes
\begin{equation}
\begin{split}
(\omega-i \partial_{z})a &=\frac{m_a^2}{2\omega}a-\frac{\g}{2}A_{\parallel}B_{0} \, ,\\
(\omega-i \partial_{z})A_{\parallel} &=\frac{\omega_{\rm pl}^2}{2\omega}A_{\parallel} -\frac{\g B_{0}}{2}a \, ,\\
(\omega-i \partial_{z})A_{\perp} &=\frac{\omega_{\rm pl}^2}{2\omega}A_{\perp} \, ,
\end{split}
\label{eq:t1}
\end{equation}
where $A_{\parallel}$ and $A_{\perp}$ are the components of the vector potential parallel and perpendicular to $\bB_{0}$. In addition we shifted the axion field as $a\rightarrow -i a$. 
Equation \eqref{eq:t1} are classical mixing equations that can be rewritten as an effective Schrödinger-like equation~\cite{Raffelt:1987im}:
\begin{equation}
i \frac{d}{dz} \Psi(z) = H\Psi(z)\, ,
\label{eq:EoM0}
\end{equation}
where the axion field and the components of the vector potential compose the `state vector'
\begin{equation}
\Psi(z) = \begin{pmatrix}
A_{\perp}\\
A_{\parallel}\\
a
\end{pmatrix}\, ,
\end{equation}
and the Hamiltonian, neglecting a part proportional to the identity, is
\begin{equation}
    H=  \begin{pmatrix}
\Delta_{\perp}& 0&0\\
0& \Delta_{\parallel}&\Delta_{a\gamma}\\
0&\Delta_{a\gamma}&\Delta_{a}
\end{pmatrix} \,,
\label{eq:mat}
\end{equation}
with the following oscillation parameters 
\begin{equation}
\begin{split}
\Delta_{a}&=-\frac{m_{a}^{2}}{2\omega}\, ,\\
\Delta_{\parallel}&=\Delta_{\perp}=-\frac{\omega_{\rm pl}^{2}}{2\omega}\, ,\\
\Delta_{a\gamma}&=\frac{g_{a\gamma} B_{0}}{2} \quad\, .\\
\end{split}
\label{eq:Deltas}
\end{equation}
Similar equations are valid in the non-relativistic case, where the Hamiltonian in eq.~\eqref{eq:mat} is multiplied by the energy-to-momentum ratio $\omega/\bar{k}$, with \begin{equation}
    \bar{k}=\sqrt{\omega^{2}-\frac{m_{a}^{2}+\omega_{\rm pl}^{2}}{2}}\,,
    \label{eq:kbar}
\end{equation}
which in general depends on $z$ because of the plasma frequency~\cite{Battye:2019aco}, this complication prevents an analytical discussion in the non-relativistic case. Note that in strongly anisotropic plasmas, it is not possible to decouple spatial derivatives of the electric field in directions perpendicular to the momentum, but a Schrödinger-like equation can still be derived for the axion-photon system~\cite{Millar:2021gzs}.

Avoided level crossings occur near `resonance points' of the axion-photon system, where $m_a=\omega_{\rm pl}$. To simplify the analysis, we take the  magnetic field direction to be constant so that the component $A_{\perp}$ decouples from the evolution, and we can focus on the two-level system formed by $a$ and $A_{\parallel}$. The relevant Hamiltonian is
\begin{equation}
H=\left(\begin{array}{cc}
\Delta_{\parallel}&\Delta_{a\gamma}\\
\Delta_{a\gamma}&\Delta_{a}\end{array}\right)\;, 
\label{eq:ham}
\end{equation}
where we (in this section) assume that the difference of oscillation parameters is given by 
\begin{equation}
\Delta_{a}-\Delta_{\parallel}=\Delta_{0}\left(\frac{z}{z_{0}}\right)\;,
\label{eq:delta}
\end{equation}
in a neighbourhood of the resonance point at $z=0$, in analogy to eq.~\eqref{eq:delta2}. It is convenient to write the first order equations derived from the Hamiltonian in eq.~\eqref{eq:ham} in terms of the axion ($a$) and photon ($A_{\parallel}$) wavefunctions
\begin{equation}
    \begin{split}
        i\,A'_{\parallel}&=\Delta_{a\gamma}\, a\, e^{-i\int_{z_{i}}^{z} dz'\,(\Delta_{a}-\Delta_{\parallel})}\;,\\
         i\,a'&=\Delta_{a\gamma}\, A_{\parallel}\, e^{i\int_{z_{i}}^{z} dz'\,(\Delta_{a}-\Delta_{\parallel})}\;,
    \end{split}
    \label{eq:EoM}
\end{equation}
where the prime represents the space derivative $\partial_{z}$ and the initial conditions  for axion-photon conversions are 
set at an arbitrary initial point $z_i$ to be:
$a(z_{i})=1$ and \mbox{$A_\parallel(z_{i})=0$}. The  two equations in eq.~\eqref{eq:EoM} can be combined to a second-order equation involving only the photon field:
\begin{equation}
    \begin{split}
        &A''_{\parallel}+\left[i(\Delta_{a}-\Delta_{\parallel})-\frac{\Delta'_{a\gamma}}{\Delta_{a\gamma}}\right]A'_{\parallel}+\Delta_{a\gamma}^{2}A_{\parallel}=0\;,\\
        &A_{\parallel}(z_{i})=0\;,\\
        &A'_{\parallel}(z_{i})=-i\Delta_{a\gamma}\;.\\
    \end{split}
    \label{eq:general}
\end{equation}
In this section, we consider only a constant magnetic field, $\Delta'_{a\gamma}=0$, but eq.~\eqref{eq:general} will be used in its full generality in subsequent sections. Moreover, note that eq.~\eqref{eq:EoM} uniquely determines the initial condition on the derivative.

It is convenient to define the rescaled photon wavefunction 
\begin{equation}
    u(z)=e^{\frac{i}{2}\int_{z_{i}}^{z}dz'(\Delta_{a}-\Delta_{\parallel})}A_{\parallel}(z)\;,
\end{equation}
which, in direct analogy with  eq.~\eqref{eq:final1}, satisfies
\begin{equation}
\begin{split}
      &u''+\left(1+\frac{1}{2n}-\frac{\tilde{z}^{2}}{4n^{2}}\right)u=0\;,\\
      &u(\tilde{z}_{i})=0\;,\\
    &u'(\tilde{z}_{i})=-i\;.\\
\end{split}
\label{eq:final}
\end{equation}
where $\tilde{z}=z\Delta_{a\gamma}$, $n=i\gamma$, the prime stands for $d/d\tilde{z}$ and
\begin{equation}
    \gamma=\frac{\Delta_{a\gamma}^{2}}{\left|\frac{d\Delta_{\parallel}}{dz}\right|}=\frac{\Delta_{a\gamma}^{2}z_{0}}{\Delta_{0}}\;,
\end{equation}
is the adiabaticity parameter for this system. A related parameter is the resonance length,
$L_{\rm res}=\sqrt{2\pi/|\Delta_{\parallel}'|}$, which is related to $\gamma$ by
\begin{equation}
 L_{\rm res}^{2} 
= \frac{\tilde L_{\rm res}^{2}}{\Delta_{a\gamma}^{2}} =   \frac{2 \pi \gamma}{\Delta_{a\gamma}^{2}} \,.
 \label{eq:Lres}
\end{equation}

The only difference between eq.~\eqref{eq:final1} and eq.~\eqref{eq:final} is the initial conditions. In eq.~\eqref{eq:final1}, we considered asymptotic initial conditions,while those of eq.~\eqref{eq:final} are set at a finite distance from the would-be level crossing. This has a technical but important consequence:  the solution, eq.~\eqref{eq:LZsol}, oscillates asymptotically  and does not have a well-defined first derivative as $w \to - \infty$, while eq.~\eqref{eq:final} sets $u'(\tilde{z}_{i})=-i$ at the initial point. 

\begin{figure}[t!]
		\vspace{0.cm}
		\centering
		\includegraphics[width=0.6\columnwidth]{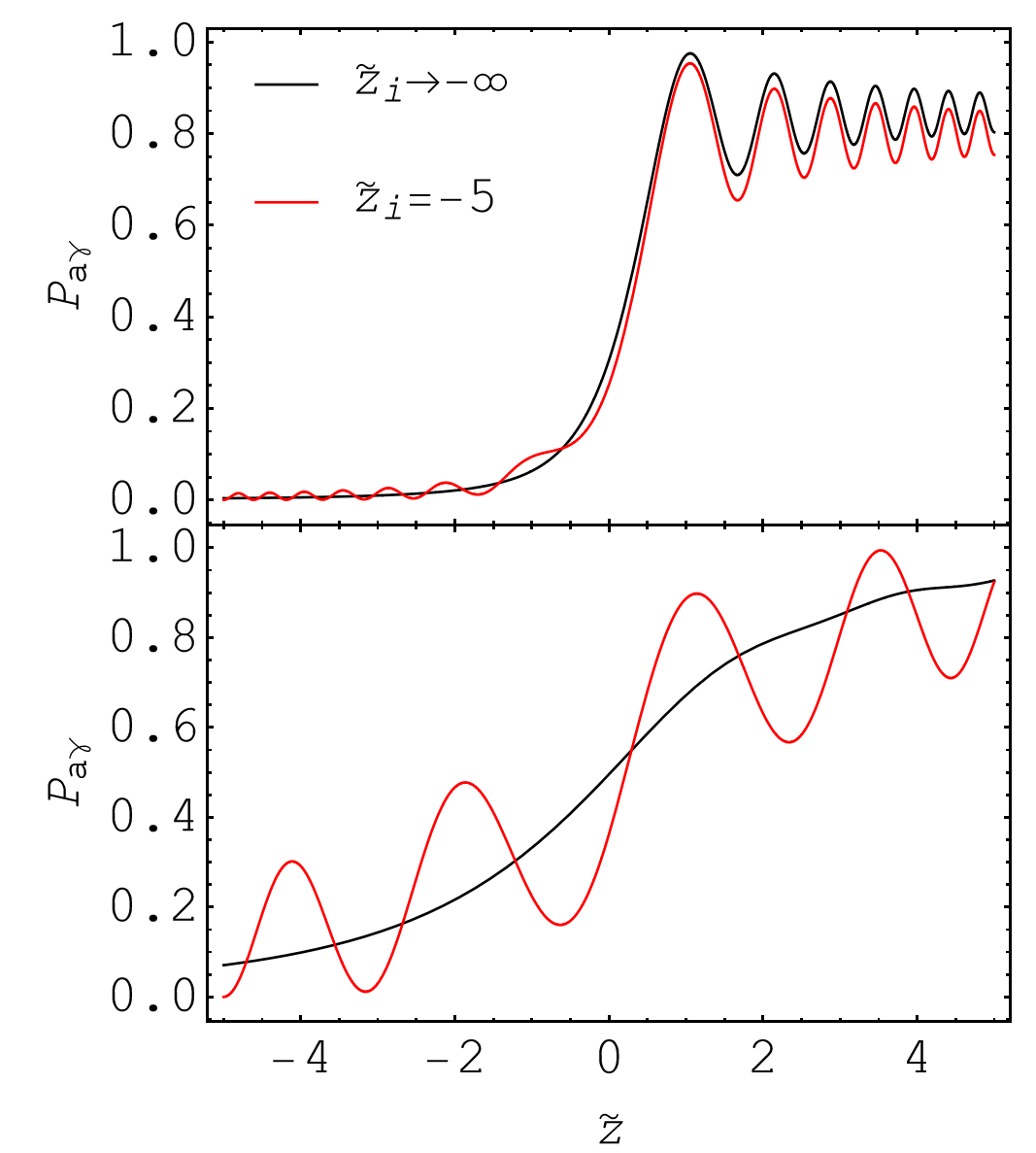}
		\caption{Comparison between the two solutions with initial condition imposed, respectively, asymptotically as discussed in section~\ref{sec:LZ} (black lines), and at a finite distance from the would-be crossing, as discussed in section~\ref{sec:LZaxion} (red lines). The upper panel shows a transition with $\gamma=0.3$ and the lower panel with $\gamma=1.5$, corresponding to resonance lengths, $\tilde L_{\rm res} = \sqrt{2\pi \gamma}$, of $1.4$ and $3.1$, respectively (see eq.~\eqref{eq:Lres}).
		} 
		\label{fig:solutions}
	\end{figure}

The solution of eq.~\eqref{eq:final} with initial conditions imposed at $\tilde z_i$ is
\begin{equation}\begin{split}
  &  u(\tilde{z})=n\frac{\mathcal{D}_{-1-n,n}\left(\frac{i\tilde{z}}{\sqrt{n}},\frac{\tilde{z}_{i}}{\sqrt{n}}\right)- \mathcal{D}_{-1-n,n}\left(\frac{i\tilde{z}_{i}}{\sqrt{n}},\frac{\tilde{z}}{\sqrt{n}}\right)}{\sqrt{n} \mathcal{D}_{-n,n}\left(\frac{i\tilde{z}_{i}}{\sqrt{n}},\frac{\tilde{z}_{i}}{\sqrt{n}}\right)+i\sqrt{n} \mathcal{D}_{-1-n,1+n}\left(\frac{i\tilde{z}_{i}}{\sqrt{n}},\frac{\tilde{z}_{i}}{\sqrt{n}}\right)-i\tilde{z}_{i} \mathcal{D}_{-1-n,n}\left(\frac{i\tilde{z}_{i}}{\sqrt{n}},\frac{\tilde{z}_{i}}{\sqrt{n}}\right)} \,,\\
   & \mathcal{D}_{p,q}\left(\frac{i\tilde{z}_{1}}{\sqrt{n}},\frac{\tilde{z}_{2}}{\sqrt{n}}\right)=D_{p}\Bigg(\frac{i\tilde{z}_{1}}{\sqrt{n}}\Bigg)D_{q}\Bigg(\frac{\tilde{z}_{2}}{\sqrt{n}}\Bigg)\,.
\end{split}
\label{eq:solgen}
\end{equation}
The conversion probability is now more complicated, and does not reduce to the familiar form ($P_{a\gamma}=1-e^{-2\pi\gamma}$).
Figure \ref{fig:solutions} shows a comparison between the (asymptotic) LZ solution  (black line) and eq.~\eqref{eq:solgen} (red line). It is apparent that the solution in eq.~\eqref{eq:solgen} oscillates not only after the crossing point, but also before, in contrast with the LZ solution in eq.~\eqref{eq:LZsol}.
The difference between the two solutions can be understood through a simple, electromagnetic analogy, as we discuss in section~\ref{sec:analogy}.

\section{Electromagnetic analogy}\label{sec:analogy}

Following the discussions in refs.~\cite{Kim:1987ss,Werner1984} we sketch an analogy between the resonant axion-photon conversion and the precession of a spin in a magnetic field. The Hamiltonian in eq.~\eqref{eq:ham}, neglecting terms proportional to the identity, can be written as
\begin{equation}
    H=\mathcal{B} \cdot \sigma\;,
\end{equation}
where $\sigma$ are the Pauli matrices and the fictitious magnetic field is 
\begin{equation}
\mathcal{B}=\begin{pmatrix}
\Delta_{a\gamma} \\ 0 \\ \frac{\Delta_{\parallel}-\Delta_{a}}{2}
\end{pmatrix}
\, .
\end{equation}
The density matrix for the axion-photon system can be decomposed in terms of the Pauli matrices as
\begin{equation}
    \rho=\frac{1}{2}\left(1+\bP\cdot\sigma\right)
\end{equation}
where we introduced the so called polarization vector 
\begin{equation}
\bP=\begin{pmatrix}2{\rm Re}\rho_{12} \\ 2{\rm Im}\rho_{12}\\ \rho_{11}-\rho_{22}\end{pmatrix}
\, .
\end{equation}
The Liouville–von Neumann equation for the density matrix (which is equivalent to the Schrödinger-like equation) translates into the evolution equation for the polarization vector in a fictitious magnetic field
\begin{equation}
    \bP'=2\mathcal{B}\times\bP\; .
\end{equation}
The photon population is explicitly given by
\begin{equation}
        p_{\gamma}=\rho_{11}=\frac{1}{2}(1+P_{z})\;.
\end{equation}

Considering axion-photon conversion, we take the initial state to be purely an axion, i.e.~$\rho_{11}=0$ and $\rho_{12}=0$.  The polarization vector is then $\bP=(0,0,-1)$,  which is anti-parallel to the $z$-axis. 

Now, if the initial condition is taken to be at $\tilde{z}_{i}\to-\infty$, then $\mathcal{B}_{z}(\tilde{z}_{i})\to \infty$ while $\mathcal{B}_{x}/\mathcal{B}_{z}\to 0$: the magnetic field becomes parallel to the $z$-axis, i.e.~it aligns with $\bP$.
By contrast, if the initial conditions are imposed at a finite point, there is always a non-vanishing $x$-component of $\mathcal{B}$, which is somewhat offset from $\bP$. 

Evolving towards the resonance point ($\tilde{z}=0$), $\mathcal{B}$ rotates in the $xz$-plane and asymptotically reverses its orientation as $\tilde z \to \infty$. During this process the polarization vector tries to follow the evolution of $\mathcal{B}$. For initially aligned vectors (read, asymptotic boundary conditions), this evolution occurs largely without precession, but when the vectors are off-set (read,  finite $\tilde z_i$), precession is unavoidable and present throughout the evolution. 
This precession effect is precisely the oscillatory behaviour shown in fig.~\ref{fig:solutions}.

\section{Regime of validity of the LZ formula}\label{sec:discussion}

In this section, we examine the conditions under which the LZ formula is applicable. Specifically, we compare its predictions to the general result in eq.~\eqref{eq:solgen}, and demonstrate that it is valid as long as the spatial extent of the resonance is much smaller than other relevant length scales and the boundaries are located far away from the resonance region. These requirements are frequently met in realistic scenarios where axion-photon conversions occur.

	\begin{figure}[t!]
		\centering
		\vspace{0.cm}
		\includegraphics[width=0.8\columnwidth]{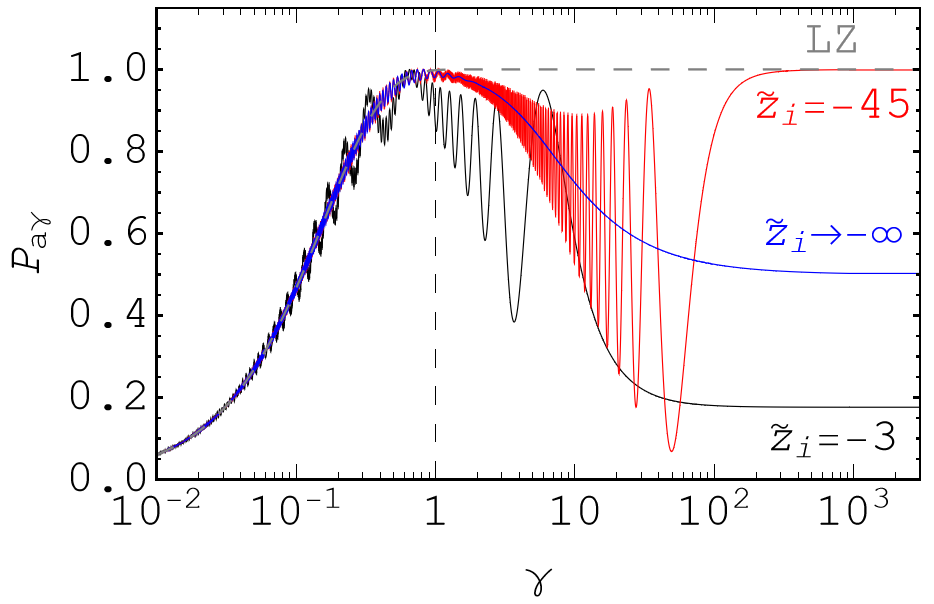}
		\caption{Conversion probability at $\tilde{z}_{f}=10$ for different adiabaticity parameters, for finite boundaries at $\tilde{z}_{i}=-3$ (black line), $\tilde{z}_{i}=-45$ (red line), and infinite boundary case (see eq.~\eqref{eq:LZsol}), $\tilde{z}_{i}\to - \infty$ (blue line). For comparison, we also show the LZ formula in eq.~\eqref{eq:LZ1} (grey dashed line), which corresponds to the conversion probability in the limit $\tilde{z}_{f}\to\infty$.} 
		\label{fig:probvsgamma}
	\end{figure}

Upon examination of the generalised LZ solution in eq.~\eqref{eq:solgen}, we find that the effect of finite boundaries on the LZ formula becomes increasingly pronounced as the adiabaticity parameter becomes large, $\gamma\gtrsim1$, and the length of the resonance region becomes similar to the length of the boundaries.
Figure \ref{fig:probvsgamma} illustrates the conversion probability at a final coordinate, $\tilde{z}_{f}=10$, as a function of the adiabaticity parameter for various initial conditions.  The black and red lines are obtained by the generalised LZ formula (eq.~\eqref{eq:solgen}) by taking initial conditions at, respectively, $\tilde z_i=-3$ and $\tilde z_i=-45$. In these cases, the conversion probability oscillates as a function of the (real-space) final coordinate even before the would-be level crossing (cf.~fig.~\ref{fig:solutions}: this is the analogue of precession discussed in section~\ref{sec:analogy}.  In addition, the finite-interval conversion probabilities are oscillatory as a function of $\gamma$, as shown in fig.~\ref{fig:probvsgamma}. The amplitudes of the oscillations increase with $\gamma$ until $\tilde L_{\rm res}(\gamma)\approx \tilde z_i$, after which they asymptote to a final, initial-condition-dependent value. 
For comparison, the blue line in fig.~\ref{fig:probvsgamma} is the result of eq.~\eqref{eq:LZsol}, i.e.~taking LZ asymptotic initial conditions at $\tilde z_i \to - \infty$ and evaluating the probability at $\tilde{z}_{f}=10$. The grey, dashed line corresponds to the LZ formula ($\tilde z_i \to - \infty$ and $\tilde z _{f}\to  \infty$). 

To better understand the previously described behaviour, we can derive simplified expressions for the conversion probabilities in the respective limits of $\gamma\gg 1$ and $\gamma \ll 1$, which we will now discuss.
In the non-adiabatic regime ($\gamma\ll1$), the constant, $\gamma$-independent term in eq.~\eqref{eq:final} becomes negligible, and the equation of motion reduces to
\begin{equation}
 \begin{split}
      &u''+\left(\frac{1}{2i\gamma}+\frac{\tilde{z}^{2}}{4\gamma^{2}}\right)u=0\;,
\end{split}   
\end{equation}
which is solved by
\begin{equation}
        u(\tilde{z})=-{\rm exp}\left[ i \left(\frac{\tilde{z}^{2}+\tilde{z}_{i}^{2}}{4\gamma} +\frac{\pi}{4}\right) \right]
        \sqrt{\frac{\gamma\pi}{2}}\left[{\rm Erf}\Bigg(\tilde{z}\frac{e^{i\frac{\pi}{4}}}{\sqrt{2\gamma}}\Bigg)-{\rm Erf}\Bigg(\tilde{z}_{i}\frac{e^{i\frac{\pi}{4}}}{\sqrt{2\gamma}}\Bigg)\right]\;,
       \label{eq:appr1}
\end{equation}
where ${\rm Erf}(z)=\frac{2}{\sqrt{\pi}}\int_{0}^{z}e^{-t^{2}}dt$. The modulus squared of eq.~\eqref{eq:appr1} represents the resonant conversion probability in the small (perturbative) mixing regime, a result also obtained in ref.~\cite{Marsh:2021ajy} with a different formalism. The asymptotic LZ formula in the non-adiabatic limit ($\gamma\ll 1$) is recovered when
\beq
\text{min}\left(z_i^2,  z_{f}^2 \right) \gg L_{\rm res}^2 \, .
\eeq

For $\gamma \gg \text{max}(\tilde{z}_i^2, \tilde{z}_{f}^2)/2$, corresponding to $L_{\rm res}^{2} \gg \text{max}(z_i^2, z_{f}^2)$, the adiabaticity parameter is large enough to make the  the term depending on $\tilde{z}$ in eq.~\eqref{eq:final} negligible. In this limit, we can solve the approximated equation
\begin{equation}
 \begin{split}
      &u''+\left(1-\frac{i}{2\gamma}\right)u=0\;,
\end{split}   
\end{equation}
which gives
\begin{equation}
  u(\tilde{z})=-i\frac{\sin\Bigg(\sqrt{1-\frac{i}{2\gamma}}(\tilde{z}-\tilde{z}_{i})\Bigg)}{\sqrt{1-\frac{i}{2\gamma}}}\;.
         \label{eq:appr2}
\end{equation}
For sufficiently large $\gamma$, this can be approximated as
\begin{equation}
        u(\tilde{z})=-i\sin(\tilde{z}-\tilde{z}_{i})\; ,
        \label{eq:appr3}
\end{equation}
which oscillates  between $0$ and $1$ as a function of the width of the integration interval. The conversion probability in this limit is highly sensitive to the  extent of the integration interval. This is the regime shown in the large $\gamma$ limit of fig.~\ref{fig:probvsgamma}, where the conversion probability converges to a constant value that strongly depends on the $\tilde{z}_{i}$.
In the case that $\text{max}(\tilde{z}_i^2, \tilde{z}_{f}^2)/2\gg 1$, there is an intermediate regime for $1\ll\gamma\ll \text{max}(\tilde{z}_i^2, \tilde{z}_{f}^2)/2$ where we have not been able to find simple analytic expressions. 

The LZ formula (eq.~\eqref{eq:LZ1}) and its generalisation (eq.~\eqref{eq:solgen}) are both derived under the restrictive assumptions of a linearly varying oscillation parameter (cf.~eq.~\eqref{eq:delta}) and a constant magnetic field. In realistic environments, both these assumptions are, at the most, approximate. We now denote the length scale over which these assumptions hold by $\lambda$, which we associate with the scale of variation of the plasma frequency: $z_{0}=\lambda$ in eq.~\eqref{eq:delta}. We expect that, in many physically interesting scenarios, the length scale for the variation of the magnetic field and the plasma frequency are comparable.

		\begin{figure}[t!]
		\vspace{0.cm}
		\centering
		\includegraphics[width=0.45\columnwidth]{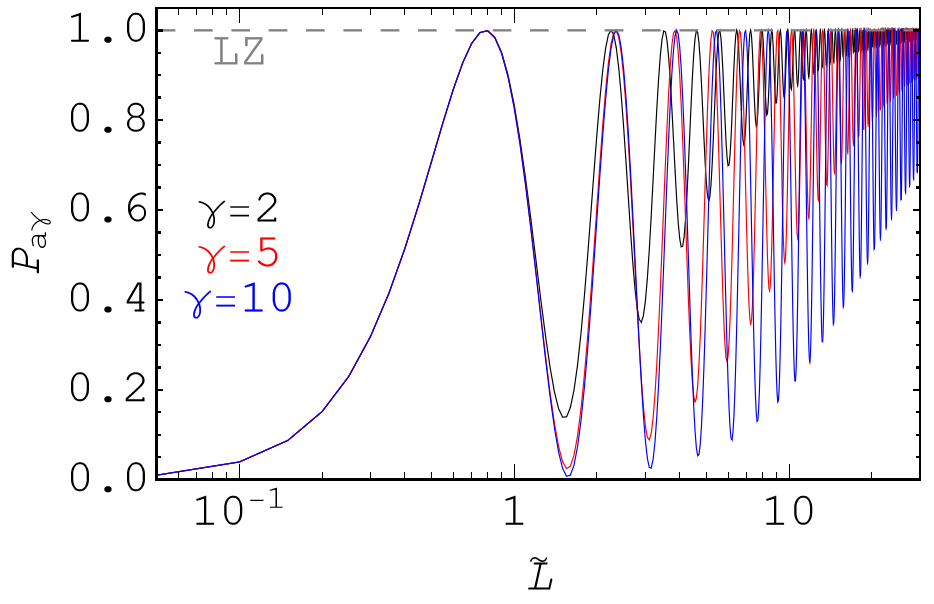}
			\includegraphics[width=0.45\columnwidth]{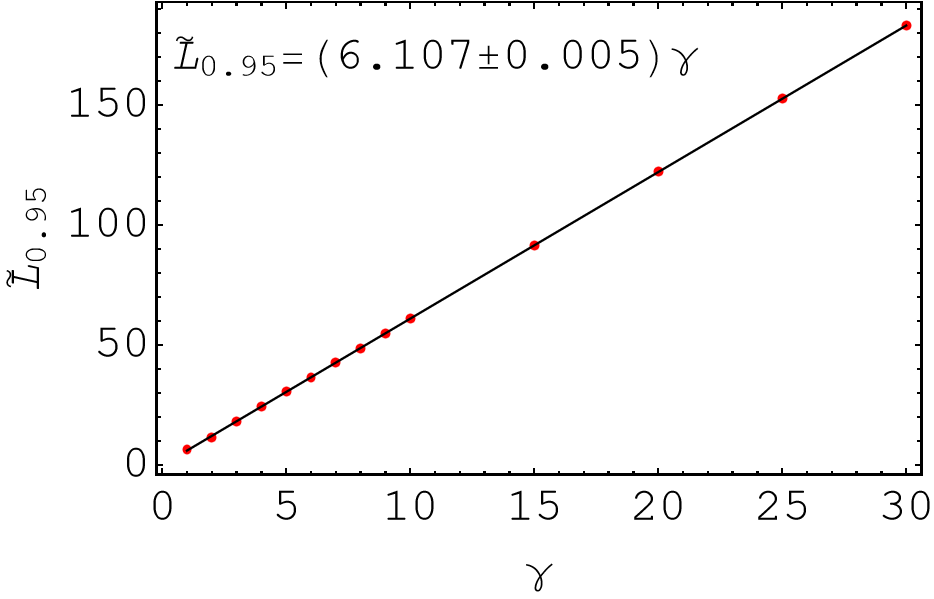}
		\caption{Left panel: conversion probability at $\tilde{L}=z_{f}$ for finite boundaries at $\tilde{z}_{i}=-\tilde{L}$, for different adiabaticity parameters $\gamma=2$ (black line), $\gamma=5$ (red line) and $\gamma=10$ (blue line). For comparison, we also show the LZ formula in eq.~\eqref{eq:LZ1} (grey dashed line), which corresponds to the conversion probability in the limit $\tilde{z}_{f}=-\tilde{z}_{i}\to\infty$. \\Right panel: evaluated points (red) and fit for $\tilde{L}_{0.95}$, as defined in the text, as function of the adiabaticity parameter.} 
		\label{fig:probvsL}
	\end{figure}
	
The LZ formula in eq.~\eqref{eq:LZ1} can be used to accurately predict the probability of non-adiabatic ($\gamma< 1$) axion-photon resonant conversions, as can be seen in fig.~\ref{fig:probvsgamma} where the LZ formula and its generalisation yield comparable results in this regime. In the perturbative regime ($\gamma \ll 1$) the predictions are in agreement with the perturbative results, as previously discussed in \cite{Battye:2019aco, Marsh:2021ajy}.
Expressed using ALP parameters, the non-adiabatic regime is determined by the condition 
\begin{equation}
   \gamma\simeq3\left(\frac{g_{a\gamma}}{10^{-11}\GeV^{-1}}\right)^{2}\left(\frac{B}{\rm G}\right)^{2}\left(\frac{m_{a}}{\rm\mu\eV}\right)^{-2}\left(\frac{\omega}{\GeV}\right)\left(\frac{\lambda}{\rm pc}\right)<1\,.
   \label{eq:crit}
\end{equation}
By contrast, in the adiabatic regime ($\gamma > 1)$ the LZ formula, despite being non-perturbative, is not always applicable. Indeed, the convergence of the result in eq.~\eqref{eq:solgen} to the LZ formula is surprisingly slow. Compare, for example, the red curve of fig.~\ref{fig:probvsgamma} to the dashed grey line: 
for $1<\gamma<50$, the system is in the adiabatic regime with an oscillation length smaller than ${\rm min}(\tilde z_i^2, \tilde z_f^2)/2$, but the full solution (red line) deviates significantly from the  LZ formula (grey line). Moreover, the left panel of fig.~\ref{fig:probvsL} shows the conversion probability given by eq.~\eqref{eq:solgen}, in the adiabatic regime as function of the width of the integration interval: $\tilde{L}=-\tilde{z}_{i}=\tilde{z}_{f}$. The three example lines convergence towards the LZ formula (that, in the figure, we consider to be exactly $1$ for $\gamma\gtrsim2$), but slowly compared to the parameter constraint  $\tilde{L}\gg 2 \sqrt{\gamma}$, discussed above in this section. In order to quantify the regime of validity of the LZ formula in the adiabatic limit, we define
\begin{equation}
    \langle P_{a\gamma}(\tilde{L})\rangle=\frac{1}{2f\tilde{L}}\int_{(1-f)\tilde{L}}^{(1+f)\tilde{L}}|u(\tilde{z})|^{2}\,d\tilde{z}\,,
\end{equation}
where $u(\tilde{z})$ is given by eq.~\eqref{eq:solgen} with $\tilde{z}_{i}=-\tilde{L}$ and $f<1$ is an arbitrary parameter to set the interval over which the oscillatory behaviour of the probability has to be averaged.  Then we introduce the length-scale $\tilde{L}_{0.95}$ such that
$\langle P_{a\gamma}(\tilde{L}_{0.95})\rangle=0.95$, requiring that the generalised LZ formula is reasonably close to the prediction of the LZ formula in the adiabatic limit ($P_{a\gamma}\simeq1$). The results of our numerical calculation of $\tilde{L}_{0.95}$ (for $f=0.2$, but the same conclusions are reached for different, sufficiently large, $f$) are shown in the right panel of fig.~\ref{fig:probvsL} as the red points, which are in remarkably good agreement with a linear fit in the form $\tilde{L}_{0.95}=6.107\gamma$ up to $\gamma=30$. For larger $\gamma$ the numerical evaluation quickly loses precision because of the fast oscillations in the conversion probability, therefore we cannot exclude a deviation from this behaviour for higher values of $\gamma$. However, extrapolating from this regime, we conclude that the LZ formula is not applicable when $\text{max}(\tilde{z}_{i}, \tilde{z}_{f})\lesssim6\gamma$ in the adiabatic limit.
This indicates that the resonant conversion probability is more sensitive to finite-distance boundary conditions than a naive analysis might suggest. 

In the next section, we present an alternative method for addressing cases with non-trivial variations in plasma frequency or magnetic field. It is especially useful in some situations where the LZ formula is not applicable, particularly when the variations of plasma inhomogeneities are slow and the axion-photon conversion is adiabatic.
	
\section{The WKB approximation}\label{sec:WKB}

If the magnetic field varies significantly, or the plasma frequency varies non-linearly, on scales comparable to the resonance length, $L_{\rm res}$, the LZ solution is no longer applicable. However, as we show in this section, a semi-classical approximation enables analytical solutions when the variation of the plasma frequency and magnetic field is slow.

We first consider the special case of degenerate states, $\Delta_{a}=\Delta_{\parallel}$, in the whole solution range. Solving eq.~\eqref{eq:final} then gives
\begin{equation}
    A_{\parallel}(z)=-i\sin\left[\int_{z_{i}}^{z}dy\,\Delta_{a\gamma}(y)\right]\;,
    \label{eq:sinsol}
\end{equation}
for a arbitrary magnetic field configuration. Note the similarity to eq.~\eqref{eq:appr3}, previously derived. The result in eq.~\eqref{eq:sinsol} is applicable when the resonance length is infinite, a regime where the LZ formula is not valid. This example of degenerate states can be of practical importance in actual physical environments, specifically in the context of axion-photon conversions in a three-dimensional plasma. In such a scenario, it is possible to identify regions, as opposed to isolated points, where the resonant condition is satisfied. One intriguing application of this concept might be in the context of axion conversion in the magnetosphere of a neutron star~\cite{Pshirkov:2007st,Fortin:2018ehg,Huang:2018lxq,Hook:2018iia,Safdi:2018oeu,Battye:2019aco,Leroy:2019ghm,Buckley:2020fmh,Foster:2020pgt,Fortin:2021sst,Witte:2021arp,Battye:2021xvt,Millar:2021gzs,Foster:2022fxn,Noordhuis:2022ljw, Witte:2022cjj}. 

		\begin{figure}[t!]
		\vspace{0.cm}
		\centering
		\includegraphics[width=0.6\columnwidth]{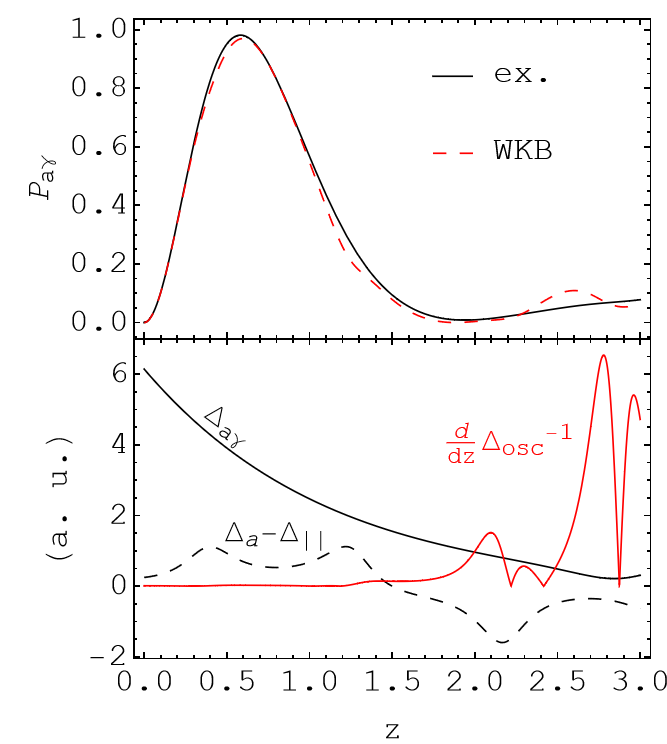}
		\caption{Application of eq.~\eqref{eq:gensolWKB} (red dashed line) and comparison with the exact numerical solution (black solid line) in the upper panel, for the magnetic field and plasma frequency shown in the lower panel (in arbitrary units). The two solutions differ when $|d\Delta_{\rm osc}^{-1}/dz|$ becomes large.} 
		\label{fig:complete}
	\end{figure}

In the more general case of a varying plasma frequency, eq.~\eqref{eq:general} cannot be solved exactly and we want to find a solution by means of the Wentzel–Kramers–Brillouin (WKB), or `semi-classical', approximation~\cite{Landau:1991wop}. For this purpose we define the photon wavefunction as $A_{\parallel}(z)=e^{i\sigma(z)}$ and eq.~\eqref{eq:general} reads
\begin{equation}
    i\sigma''-(\sigma')^{2}+(\Delta_{\parallel}-\Delta_{a})\sigma'+\Delta_{a\gamma}^{2}=0\;,
    \label{eq:WKB}
\end{equation}
where we, for now, have assumed a constant magnetic field. The WKB approximation requires that
\begin{equation}
    \Big|\frac{\sigma''}{(\sigma')^{2}}\Big|=\Big|\frac{d}{dz}\left(\frac{1}{\sigma'}\right)\Big|=\Big|\frac{d\Delta_{\rm osc}^{-1}}{dz}\Big|\ll1\;,
    \label{eq:wkbass}
\end{equation}
where $\Delta_{\rm osc}=\sqrt{(\Delta_{a}-\Delta_{\parallel})^{2}+4\Delta_{a\gamma}^{2}}$.
 Equation \eqref{eq:WKB} now becomes, approximately,  a quadratic equation in $\sigma'$
\begin{equation}
   -(\sigma')^{2}+(\Delta_{\parallel}-\Delta_{a})\sigma'+\Delta_{a\gamma}^{2}=0\, .
\end{equation}
The general solution of this equation  is a  linear combination of 
\begin{equation}
    \begin{split}
        \sigma^{(\pm)}(z)&=\frac{1}{2}\int_{z_{i}}^{z} dy\left(\Delta_{\parallel}(y)-\Delta_{a}\pm\Delta_{\rm osc}(y)\right)\, .
    \end{split}
\end{equation}
For the initial conditions given by eq.~\eqref{eq:general}, the solution is given by
\begin{equation}
    A_{\parallel}(z)=-i\frac{2\Delta_{a\gamma}}{\Delta_{\rm osc}}\sin\left[\frac{1}{2}\int_{z_{i}}^{z}dy\,\Delta_{\rm osc}(y)\right]\;.
     \label{eq:gensolWKB}
\end{equation}
Equation~\eqref{eq:gensolWKB} is derived in the assumption that the magnetic field is constant and the plasma frequency varies slowly (cf.~eq.~\eqref{eq:wkbass}). For $\Delta_{a}=\Delta_{\parallel}$, we recover precisely eq.~\eqref{eq:sinsol}, which applies to arbitrary magnetic fields. Consistently with these observations, we assume that eq.~\eqref{eq:gensolWKB} applies to any configuration of magnetic field and plasma frequency that satisfy the WKB conditions in eq.~\eqref{eq:wkbass}.  Figure~\ref{fig:complete} demonstrates the use of the WKB approximation in case varying magnetic field and plasma frequency.
Our conclusion is that the WKB solution is found to be a good approximation when the oscillation length varies slowly enough that eq.~\eqref{eq:wkbass} is satisfied, as can be verified from the lower panel of fig.~\ref{fig:complete}.

\section{Conclusions}\label{sec:conclusions}
	
We have investigated the LZ formula's validity for calculating the axion-photon resonant conversion probability. The LZ formula is a widely used tool in axion physics, applied in various astrophysical and cosmological contexts; however, this simple formula does not apply in some physical situations. The LZ formula relies on two assumptions: 1) the coherence length of the magnetic field is much larger than the length over which the plasma frequency varies, and 2) the resonance length is shorter than the scale of plasma variation. In a realistic environment, the magnetic field and plasma frequency inhomogeneities are related and have similar characteristic length scales, violating the first assumption. Additionally, the resonance length may be larger than the plasma length scale, violating the second assumption.

To better understand the impact of plasma inhomogeneities on conversion probability, we generalised the LZ formula to allow for an arbitrary location of the boundary, enabling us to model plasmas with arbitrary coherence lengths. We found that the LZ formula is not accurate when the boundary is too close to the region of conversion, as determined by the resonance length, particularly when the conversion is adiabatic (i.e., $\gamma\gtrsim1$).

Our findings confirm that the LZ formula is applicable for conversions of dark matter axions into radio waves in the magnetosphere of neutron stars~\cite{Battye:2019aco,Foster:2022fxn}, as well as in the case of conversions in the primordial magnetic field of the early universe, which may cause distortions in the cosmic microwave background~\cite{Mirizzi:2009nq,Mukherjee:2018oeb}. This is because the low energies of the axions in these cases result in short resonance lengths, compared to the length of the plasma involved in the conversions ($\gamma\ll \text{max}(\tilde{z}_i^2, \tilde{z}_{f}^2)/2$). 
By contrast, adiabatic resonant conversions of high-energy axions might affect the photon spectra from active galactic nuclei~\cite{Hochmuth:2007hk}.\footnote{The strong limits on axions from spectral distortions of AGN spectra derived in e.g.~\cite{Berg:2016ese, Marsh:2017yvc, Reynolds:2019uqt, Reynes:2021bpe, Matthews:2022gqi} conservatively assume only non-resonant conversion, and are unaffected by these considerations.}  Our generalised LZ formula suggests that in this case, the conversion probability is not guaranteed to be maximal and may be significantly less than unity, contrary to the common belief that adiabatic conversions always result in a high conversion probability even in realistic plasmas.

As discussed, the LZ formula is often not applicable for non-trivial configurations of magnetic field and plasma frequency. To address this, we have proposed a semi-classical approximation that can be used to calculate the conversion probability in situations where the LZ formula is not applicable. For instance, when the resonance occurs over a segment rather than a single point, a scenario that may be relevant for axion-photon conversion in three dimensions. 
Our findings highlight the importance of considering alternative methods for analysing resonant axion-photon mixing and the need for further research in this area.

\section*{Acknowledgements}
PC and DM are supported by the European Research Council under Grant No. 742104 and by the Swedish Research Council (VR) under grants 2018-03641 and 2019-02337. This article is based upon work from COST Action COSMIC WISPers CA21106, supported by COST (European Cooperation in Science and Technology).

\bibliographystyle{apsrev4-1}
\bibliography{LZ}

\end{document}